\documentclass[a4paper]{article}

\usepackage{INTERSPEECH2020}
\usepackage{amssymb,amsmath,epsfig}
\usepackage{textcomp}
\usepackage{amsfonts,amsthm,amsopn}
\usepackage{adjustbox}
\usepackage{tikz}
\usepackage{bm}
\usepackage[capitalise]{cleveref}
\usepackage{graphicx,caption,lipsum}
\usepackage{booktabs,multirow}
\usepackage{placeins}
\usepackage{algpseudocode,algorithm}
\usepackage{color}
\usepackage{enumerate}   
\usepackage{adjustbox}
\usepackage[normalem]{ulem}

\newcommand*\circled[1]{\tikz[baseline=(char.base)]{
    \node[shape=circle, draw, inner sep=0.5pt, 
        minimum height=10pt] (char) {\vphantom{1g}#1};}}
        
\ninept
\def\reg{{\rm\ooalign{\hfil
      \raise.07ex\hbox{\scriptsize R}\hfil\crcr\mathhexbox20D}}}

\title{Cross-domain Adaptation with Discrepancy Minimization for  Text-independent Forensic Speaker Verification}

\makeatletter
\def\name#1{\gdef\@name{#1\\}}
\makeatother 
\name{Zhenyu Wang, Wei Xia, John H.L. Hansen{\em } }
\email{\{Zhenyu.wang, wei.xia, john.hansen\}@utdallas.edu}

\address{
 Center for Robust Speech Systems (CRSS), University of Texas at Dallas, TX 75080 
{\small \tt}}

\newcommand{\vct}[1]{\boldsymbol{\mathbf{#1}}} 

\begin{document}

\maketitle
\begin{abstract}
\vspace{-1ex}
Forensic audio analysis for speaker verification offers unique challenges due to location/scenario uncertainty and diversity mismatch between reference and naturalistic field recordings. The lack of real naturalistic forensic audio corpora with ground-truth speaker identity represents a major challenge in this field. It is also difficult to directly employ small-scale domain-specific data to train complex neural network architectures due to domain mismatch and loss in performance. Alternatively, cross-domain speaker verification for multiple acoustic environments is a challenging task which could advance research in audio forensics. In this study, we introduce a CRSS-Forensics audio dataset collected in multiple acoustic environments. We pre-train a CNN-based network using the VoxCeleb data, followed by an approach which fine-tunes part of the high-level network layers with clean speech from CRSS-Forensics. Based on this fine-tuned model, we align domain-specific distributions in the embedding space with the discrepancy loss and maximum mean discrepancy (MMD). This maintains effective performance on the clean set, while simultaneously generalizes the model to other acoustic domains. From the results, we demonstrate that diverse acoustic environments affect the speaker verification performance, and that our proposed approach of cross-domain adaptation can significantly improve the results in this scenario.
\end{abstract}


\section{Introduction}
\label{sec:intro}
The need for forensic speaker recognition arises when an individual contributes his or her voice as evidence such from telephone recordings ,wiretaps, audio surveillance, or informant recordings~\cite{hansen2015speaker}. 
The use of technology for forensic speaker recognition has been considered as early as 1926 based on speech waveform analysis~\cite{wigmore1926new}. 
Today, forensic speaker recognition is commonly performed by human expert phoneticians, fully or partially, who generally have backgrounds in linguistics and statistics. Full or assisted automatic approaches are also considered as an efficient tool for forensic speaker recognition.

Speaker verification determines whether a test utterance belongs to the target person. Previously, the segment-level vector that represents the speech signal, called i-vector, with a probabilistic linear discriminant analysis (PLDA) backend, has dominated the text-independent speaker verification research field~\cite{dehak2010front}.
Additionally, i-vector variants have been widely used in multiple fields of paralinguistic speech attribute recognition ~\cite{dehak2011language,li2014speaker}. 
With the emergence of the large speaker labeled audio datasets and growing computational resources, there is increased interest in applying more effective networks such as x-vector and other neural network architecture to speaker verification tasks~\cite{snyder2018x,snyder2016deep,li2017deep,cai2018exploring,lei2014novel}.

A fundamental assumption in machine learning is that training data and test data are drawn from the same distribution~\cite{ben2010impossibility,mansour2009domain}. 
Since this assumption can be undermined by many factors, such as deploying a well-trained model from a large-scale dataset to specific tasks with emotion, duration, language mismatch, etc.~\cite{xi2019speaker,lin2020multi,xia2018speaker,xia2019cross,vesely2016sequence}. 
In general, effective audio data set collection under multiple acoustic environments can help, where a portion of the data encompasses source/reference information, while other data can be obtained as field, unknown, or target domain with diverse acoustic mismatch ~\cite{csurka2017domain}. The source domains and target domain can be similar, but are generally not identical. In domain adaptative training, maximum mean discrepancy (MMD)~\cite{gretton2012kernel} and triplet discrepancy loss~\cite{zhu2019aligning,xu2018deep} are used for discrepancy evaluation in some domain adaptation motivated tasks~\cite{lin2018reducing,lin2018multisource}.

Speech is a highly variable phenomenon. Intrinsic speaker characteristics represent speech traits which are controlled/dependent on the speaker, vs. extrinsic characteristics which are dependent on audio capture and environmental factors~\cite{hansen2015speaker}. Intrinsic properties include the speaker’s age, sex/gender, vocal effort, emotional and physical state (e.g. angry, sad, stressed, distracted, etc.). Extrinsic properties include the recording conditions such as microphone type and placement, background noise and reverberation due to environmental scenarios. The multifaceted sources of variation pose the greatest challenge to accurately model and recognize a speaker regardless of the approaches employed.

As noted above, intrinsic and extrinsic characteristics of audio samples may affect the performance of speaker verification systems in discriminating speakers. Here, we temporarily freeze  intrinsic variation of the speaker, and place primary research interest on diversity of acoustic environments as extrinsic variation. This represents a core challenge in most forensic voice analysis investigations. In this context, there does exist shared domain-invariant information across these acoustic environments.  In this study, we employ cross-domain adaptation to learn the domain-invariant information across domains containing extrinsic variations of the speaker identity. Firstly, we use a large-scale dataset to train a common feature extractor for speaker identities. The well-trained model is fine-tuned with the clean set of our small in-domain data. Finally, we perform the proposed cross-domain adaptation with discrepancy minimization to extend the fine-tuned model to all other target sets with learned domain-specific subnets. It can significantly improve the speaker verification performance on other sets in different acoustic environments, and still keep an excellent performance on the clean set, alleviating the catastrophic forgetting problem~\cite{kirkpatrick2017overcoming}. 
Experimental results show that the speaker verification performance in a specific dataset with multiple acoustic domains is improved by a large margin using the proposed cross-domain adaptation approach.


\section{Overview of the Adaptation Framework}
\label{sec:adapt}
Since the pre-trained system is trained on a large-scale dataset, the model adaptation should be applied to alleviate the impact of data mismatch between training data and data in specific domains. This adaptation includes two stages: adaptation on the clean set and the cross-domain adaptation among different acoustic environments.

\vspace{-1ex}
\subsection{Pre-trained systems}
We use an i-vector system as one of our pre-trained systems. The other pre-trained system is based on neural networks, by employing a multiple-layer residual neural network (ResNet)~\cite{he2016deep} to build a front-end main body as a common feature extractor, followed by a learnable dictionary encoding (LDE) layer~\cite{cai2018exploring} to encode a variable-length temporal sequence into a fixed-dimensional representation at the utterance level.

\vspace{-1ex}
\subsection{Adaptation to clean data}
The first-stage adaptation method of the GMM-based i-vector system is MLLR, which uses a linear transformation of Gaussian model parameters on top of the UBM~\cite{ganitkevitch2005speaker}. 

In the first step of our CNN-based system adaptation, the new model is initialized by weights of the pre-trained CNN-based model, freezing the first three layers of ResNet, then it continues to be trained with clean data. The learning rate of this adaptation is 10x smaller than that used in the training process. The inputs across domains will share the domain-independent adapted feature extractor to generate embeddings for further cross-domain adaptation, and the process through the entire common feature extractor is symbolized as $\Phi_0$.

\vspace{-1ex}
\subsection{Cross-domain adaptation}
The objective of our second-stage adaptation is to learn domain-invariant representations in the embedding space for all domains. With the fine-tuned model on the clean set in the first step, following a domain adaptation setup, we take the clean data as the ``source'' domain data. LENA-booth, far-field, and noisy data (CRSS-Forensics corpus explained in Sec. 3.1.2) are taken as the ``target'' domain data, which also have speaker labels. At first, a discrepancy loss will be used to reduce the distinction of clean sample representations across domain-specific embedding extractors, and this process is symbolized as $\Phi_1$. Next, for process $\Phi_2$, it is easier to extract domain-invariant representations for each source/target pair and then subsequently align the distributions of source and target domain afterwards. In the end, domain-specific classifiers are employed to optimize verification performance for individual domains. 

\vspace{-1ex}
\subsubsection{Cross-domain discrepancy minimization}
The embedding extractors learned from target domains are likely to make wrong decisions on clean samples near the class boundaries. Based on this fact, a discrepancy loss is employed to compute the absolute values of differences among all pairs of representations from domain-specific embedding extractors. The discrepancy loss function is formulated as,
\vspace{-1ex}
\begin{align}
\begin{split}
&Loss_{dis} = \frac{2}{N(N-1)}\times  \\
&\sum_{i=1}^{N-1} \sum_{j=i+1}^{N} \mathbb{E}_{\vct x \in \mathcal{D}_{s}} \left | \Phi_{1}^{i}(\Phi_{0}(\vct x^{s})) - \Phi_{1}^{j}(\Phi_{0}(\vct x^{s})) \right | 
\end{split}
\vspace{-1ex}
\end{align}
\noindent In Eq. 1, $\vct{x}^s$ is the input feature vector of the clean set. $\Phi_1^i$ is an intermediate process between the common feature extractor and discrepancy evaluation in domain $i$, $N$ domains are included.

\subsubsection{Pair-wise distribution alignment}
We employ MMD as a pair-wise discrepancy evaluation, which is a kernel two-sample test for the hypothesis $\mathcal{D}_{s} = \mathcal{D}_{t}$ (target and source domains are the same) based on the observed samples. Given that generated distributions are identical, MMD assumes that all the corresponding statistics are the same. The definition is given in Eq. 2, which estimates the discrepancy between each pair of source and target domain,
\begin{align}
\begin{split}
& MMD\left(\mathcal{D}_{s}, \mathcal{D}_{t}\right)_{h} \triangleq \\
& \left\| \mathbb{E}_{\mathcal{D}_{s}}[\Phi_{2}^{h}(\Phi_{1}^{h}(\Phi_{0}(\vct x^{s})))]
- \mathbb{E}_{\mathcal{D}_{t}}[\Phi_{2}^{h}(\Phi_{1}^{h}(\Phi_{0}(\vct x^{t})))] \right\|_{h}^{2}
\end{split} 
\end{align}
\noindent Where $h$ is a specific target domain, $\vct{x}^s$ and $\vct{x}^t$ are sequence input features in the source and target domains. $\Phi_2^h$ is an intermediate process between the discrepancy evaluation and MMD evaluation in domain $h$. Given $N$ target domains, $\widehat{MMD} (\mathcal{D}_s, \mathcal{D}_t)_h$ is an unbiased estimator of $MMD (\mathcal{D}_s, \mathcal{D}_t)_h$, where $\mathbb{E}_{\vct{x} \in \mathcal{D}} = 1/n \sum_{i=1}^{n}  \Phi_{2}^{h}(\Phi_{1}^{h}(\Phi_{0}(\vct{x}_i)))$. Domain-invariant representations for each paired source and target domain can be learned by minimizing the $Loss_{mmd}$,
\begin{align}
Loss_{mmd}=\sum_{t=1}^{N} \widehat{MMD} (\mathcal{D}_s, \mathcal{D}_t)_h
\end{align}

\subsubsection{Domain-specific classification}

Each domain-specific embedding extractor is followed by a softmax classifier. We add this classification loss for each classifier to ensure that the embedding distribution performance is improved in each target domain. We use a cross-entropy $\mathcal{J}$ loss as the following,
\begin{align}
Loss_{cls}=\sum_{t=1}^{N} \mathbb{E}_{\vct x \in \mathcal{D}_{t}} \mathcal{J}(\Phi_{2}^{t}(\Phi_{1}^{t}(\Phi_{0}(\vct x_{i}^{t}))), y_{i}^{t})
\end{align}
\noindent Given $N$ target domains, the classification losses are computed for each domain-specific classifiers.

\subsubsection{Cross-domain adaptation framework}

Given a sequence data input, the common feature extractor projects a sequence data into a temporal orderless embedding in a shared process $\Phi_0$. $\Phi_1$ and $\Phi_2$ processes of the cross-domain adaptation method have three independent subnets corresponding to specific target domains.
The domain-invariant representations are learned in this unshared process
$\Phi_1$ to minimize discrepancies among target domains. 
The final embeddings for new target domains are generated through the unshared process $\Phi_2$ from domain-specific classifiers. Also, distributions of both target and source samples are aligned among each domain in $\Phi_2$. Assuming the cross-domain adaptation can effectively learn a domain-invariant representation, 
we compute the clean set embedding as the average of target domain embeddings here. The framework of the cross adaptation in each step is illustrated in Fig. 1. 
The multi-task loss function is formulated as,
\begin{align}
Loss_{total} = \mu (Loss_{mmd} + Loss_{dis}) + Loss_{cls}
\end{align}
\noindent Where $\mu$ is a variant adaptation factor with a progressive schedule from 0 to 1. in order to stabilize parameter sensitivity in the early adaptation stage.

\vspace{-1ex}
\begin{figure}[tbp]
  \centering
  \includegraphics[width=0.99\linewidth, height=8cm]{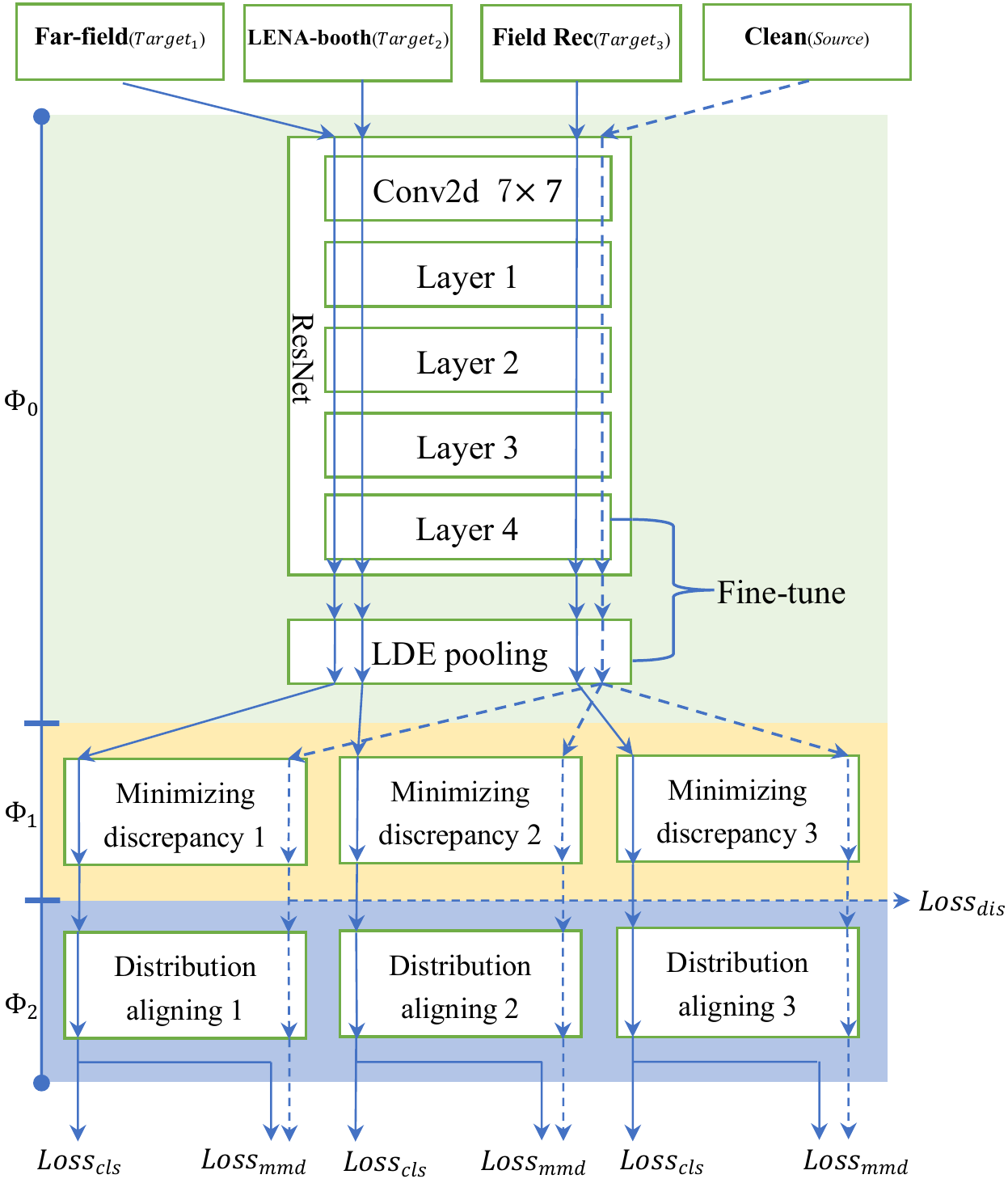}
  \caption{Cross-domain adaptation framework.}
  \label{fig:se2}
  \vspace{-3.5ex}
\end{figure}

\section{Experiments}

\subsection{Data description}
\vspace{-1ex}
\subsubsection{VoxCeleb}
We use VoxCeleb 1 and 2~\cite{nagrani2017voxceleb} dataset which is extracted from videos on YouTube as the training data for our pre-trained systems. Videos included in the dataset are shot in a large number of challenging visual and auditory environments including background chatting, laughter, overlapping speech, and varying room acoustics~\cite{chung2018voxceleb2}. 
The training dataset contains over 1,261,189 utterances from 7273 identities. 

\vspace{-1ex}
\subsubsection{CRSS-Forensics}
The CRSS-Forensics corpus\footnote{The CRSS-Forensics corpus will be released with a license.} contains read speech, prompted speech, and spontaneous speech in three conditions: (i) clean (sound booth), (ii) noisy (field recordings), and (iii) high stress (actual police interviews).  Two phases are included in the recording process. Phase-1 contains speech data recorded in different acoustic environments such as a sound booth and public environments. Speech in Phase-2 is collected in a law enforcement facility, using an interview room with an actual police officer/detective. Figure 2 shows sample recording environments for the sound booth, public environments, and police interview room.

For the sound booth of Phase-1, speech data is recorded using microphones (sample rate: 44.1 kHz) and a participant body-worn mobile data collection platform called LENA unit (sample rate: 16 kHz). Microphones are positioned at 4 different locations in the booth; the distances from each microphone to the speaker are 10 inches and 0.8, 4, and 8 feet, as shown in Fig. 2 (a).In field environments, speech is collected by the LENA unit worn by the participant, and the recording environments include seven indoor and outdoor places such as hallway (Fig.2 (b)), cafeteria (Fig.2 (c)), walking path (Fig.2 (d)), parking lot (Fig.2 (e)), etc.
In Phase-2, a detective interviews the participant concerning a specific investigative scenario while following standard procedures in an interview room (Fig. 2 (f)). 
\begin{table}[tbp]
  \centering
  \renewcommand{\arraystretch}{1.1}
  \caption{Data statistics for the CRSS-Forensics corpus}
  \vspace{-2ex}
    \begin{tabular}{|c|c|c|c|c|c|}
    \hline
    \multicolumn{1}{|c|}{Phase} & Session type & Duration & \multicolumn{1}{c|}{Spk} & \multicolumn{1}{c|}{Utt} \\
    \hline 
    
    \multirow{4}{*}{1} & Clean & 64 h  & 75    & 6788 \\ 
        \cline{2-5}
         & Far-field & 64 h  & 75    & 6788 \\
        \cline{2-5}
         & LENA-booth & 33.9 h & 75    & 4060 \\
        \cline{2-5}
          & Field Recording & 99.4 h & 75    & 11965 \\
    \hline
    2     & Police Interview & 40.8 h & 58    & 4615  \\
    \hline
    \end{tabular}%
  \vspace{-6ex}
\end{table}%

Table 1 summarizes the specific data size for each session. For the 75 speakers in the corpus, there are 65 English native speakers and 10 non-native speakers, with 27 male speakers, and 48 female speakers. Each participant was allowed to opt-out of Phase-2 (i.e. IRB protocol due to high stress level exposure), so there are 17 speakers fewer from the police interview set. 

In this current study, data from Phase-1 is used for cross-domain adaptation. We note that various recording environments are considered as the extrinsic characteristics for audio samples, while speech from speakers under stress in police interviews are intrinsic variations. Consequently, data in Phase-2 is not compatible with data in Phase-1 for domain-invariant information extraction. We split data in Phase-1 at a ratio of 7:1:2 for training, enrollment, and test data. The duration of each speech record is segmented into 30 seconds. The number of test trials in each set is 47684, 47684, 24564, 47778 for the clean, far-field, LENA-booth, and field-recording, respectively. For sound booth data, speech data with mics close to the speaker (10 inches \& 0.8 feet) are viewed as clean in the target domain for first-step adaptation, and speech data from the remaining two distant mics are viewed as far-field data. Data captured by the LENA body-worn unit was used to explore channel mismatch influence.
\vspace{-2ex}
\begin{figure}[hbtp]
  \centering
  \includegraphics[width=0.95\linewidth, height=3cm]{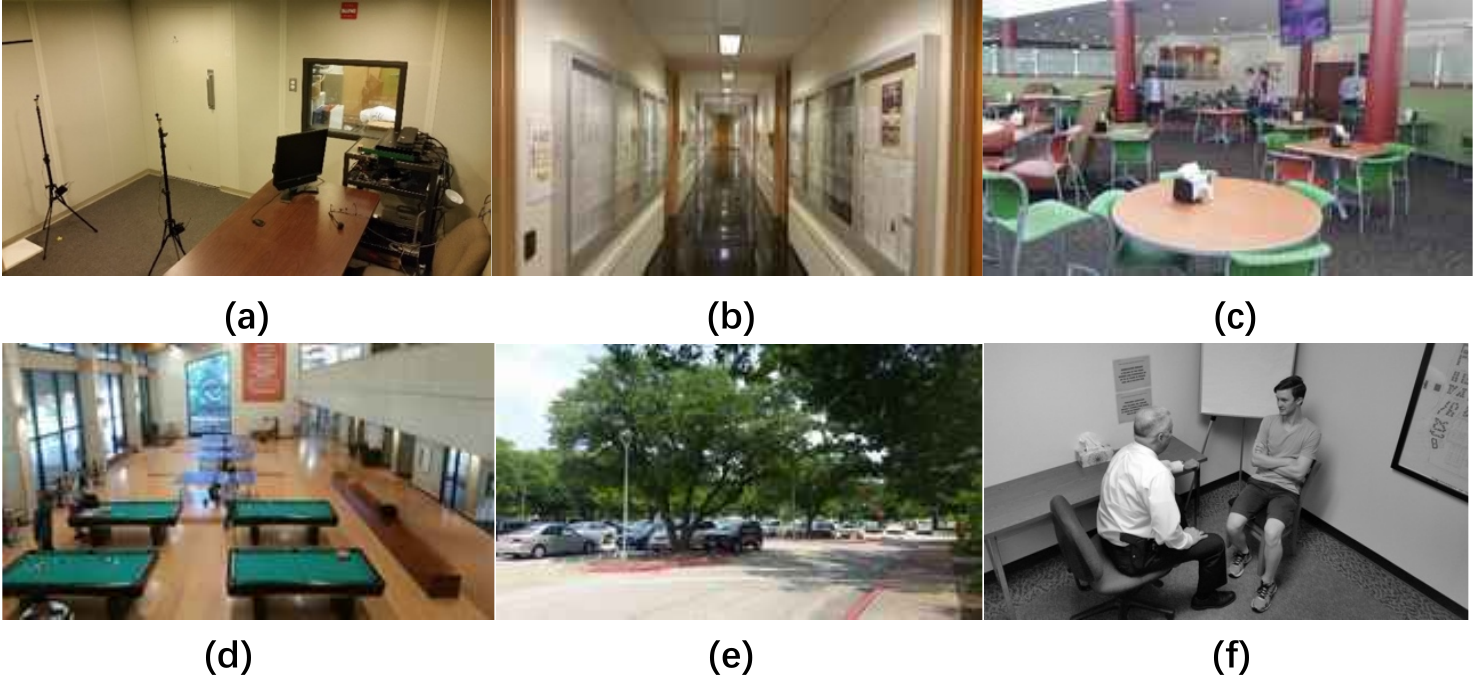}
  \vspace{-2ex}
  \caption{Forensic recording environments - (a) sound booth, (b-e) field recording, (f) police interview.}
\end{figure}
\vspace{-4ex}

\vspace{-1ex}
\subsection{i-vector system configuration}
Gender independent i-vector extractors were trained on the VoxCeleb 1 and 2 to produce the 400-dimensional i-vectors. 20-dimensional MFCCs were augmented with their delta and double-delta coefficients, composing 60-dimensional MFCC feature vectors.
MLLR was used for adapting the basic i-vector model to the clean data.

\vspace{-1ex}
\subsection{CNN-based system configuration} 
A 30-dimensional Fbank was extracted with a frame length of 25 ms and frame shift of 10 ms as the acoustic feature for each utterance. Kaldi energy-based VAD was also applied to remove the silent segments in each utterance. Then the raw acoustic feature was input into a widely-used ResNet-34 architecture with Angular-Softmax~\cite{liu2017sphereface}. The corresponding channel sizes for conv1, layer1, layer2, layer3, and layer4 were 16, 16, 32, 64, and 128 respectively. The number of dictionary components in the LDE layer was 64. 
Adam optimizer was employed for the network training with betas of (0.9, 0.98), weight decay of 1e-4, and the AMSGrad variant. The learning rate was adjusted with the warm-up scheduling named “Noam” in~\cite{vaswani2017attention}.

In the fine-tuning stage, each input data was cropped to 1200 frames, which approximated the mean length of an utterance after VAD. The parameters in layer 4 and the LDE layer were updated, and parameters for all other layers were frozen.

\vspace{-2ex}
\subsection{Cross-domain adaptation configuration}
We did cross-domain adaptive training for layer 4 in ResNet, LDE layer, and the cross-domain adaptation block via back propagation. The cross-domain subnet of each domain was composed of four fully-connected layers: (4096,1024), (1024,1024), (1024,256), and (256,256); the first two and last two layers were applied for minimizing discrepancy and distribution alignment, respectively. Since the cross-domain adaptation block was trained from scratch, we set its learning rate to be 10x larger than that of the layer 4 in ResNet and LDE layers. The learning rate of the cross-domain adaptation block was adjusted using the formula,
\vspace{-1ex}
\begin{align}
\eta_p = \frac{\eta_0}{(1+ \alpha p)^{\beta}}
\end{align}
\noindent Where $\eta_0 = 0.01$, $\alpha=10$, $\beta=0.75$ and $p$ linearly changes from 0 to 1 corresponding to the training steps. Since there exist no parameter-wise differences between each subnet in the early adaptation stage, $Loss_{mmd}, Loss_{dis}$ may result in noisy activations. To stabilize parameter sensitivity, a progress strategy~\cite{ganin2014unsupervised} is used for $Loss_{total}$:
\vspace{-2ex}
\begin{align}
\mu = \frac{2}{exp(-\theta p)} - 1
\end{align}
\vspace{-1ex}
Where $\theta=10$ makes $\mu$ change from 0 to 1.

\section{Results and Discussion}
We compare a GMM-based i-vector system with a system based on neural networks for the speaker verification task. Results of the pre-trained model, the fine-tuning step, and the cross-domain adaptation (CDA) process are presented respectively for the two systems. In the cross-domain setup, there are channel, reverberation, and noise mismatches between datasets in various domains, which can be used to analyze how extrinsic variation in multiple acoustic environments impacts the speaker verification performance. 
Practically, it is very inefficient and computational expensive to maintain multiple models for different domains. 
The cross-domain adaptation method can help us train a single model to keep an excellent performance on the source-domain clean set and improve the verification result of the fine-tuned model on each target domain by a large margin. Table 2 shows systems’ testing results in terms of equal error rate (EER) for each set. Due to the small absolute values shown in each set, we use a relative decrease (RD\%) of EER to evaluate system improvement in each step. MLLR and fine-tuning are used by i-vector and CNN-based systems respectively in the clean set adaptation step.

Before any adaptation is applied, embeddings extracted from the raw common feature extractor are employed for speaker verification. Two systems show good results with low EER in general. Testing results for diverse domains vary a lot for two systems. The noisy data collected by the LENA unit in public environments shows the worst testing performance. Clean data and LENA-booth data are both collected in the sound booth,where the former one is recorded by microphones and achieves a better result than the latter recorded with the LENA unit. This indicates that the noise and channel mismatches considerably affect the speaker verification performance. The far-field condition generally has less of an influence on speaker verification. Far-field data in our corpus is recorded by mics in a sound booth, where the voice is less likely to be distorted by the given distance as it travels in the wild. 
\vspace{-2ex}
\begin{table}[htbp]
  \centering
  \renewcommand{\arraystretch}{1.4}
  \caption{Equal error rate of systems in various domains}
  \vspace{-2ex}
    \begin{adjustbox}{width=1.\columnwidth, center}
    \begin{tabular}{|l|c|c|c|c|}
    \hline
    \multicolumn{1}{|l|}{} & \multicolumn{1}{c|}{Clean} & \multicolumn{1}{c|}{LENA-booth} & \multicolumn{1}{c|}{Far-field} & \multicolumn{1}{c|}{Noisy} \\
    \hline
    \circled{1} Raw i-vector & 0.58\% & 0.83\% & 0.69\% & 2.73\% \\
    \hline
    \circled{2} i-vector MLLR & 0.22\% & 0.73\% & 0.34\% & 2.37\% \\
    \hline
    $\circled{1} \rightarrow \circled{2}$ RD & $\uline{62.07\%} \downarrow$ & $\uline{12.05\%} \downarrow$ & $\uline{50.72\%} \downarrow$ & $\uline{13.19\%} \downarrow$ \\
    \hline
    \circled{3} i-vector CDA & 0.21\% & \textbf{0.63\%} & 0.23\% & 2.28\% \\
    \hline
    $\circled{2} \rightarrow \circled{3}$ RD & $\uline{4.55\%} \downarrow$ & $\uline{13.70\%} \downarrow$ & $\uline{32.35\%} \downarrow$ & $\uline{3.80\%} \downarrow$\\
    \hline
    \hline
    \circled{4} Raw CNN & 1.15\% & 1.83\% & 1.03\% & 2.43\% \\
    \hline
    \circled{5} CNN fine-tune & 0.113\% & 1.49\% & 0.20\% & 2.06\% \\
    \hline
   $\circled{4}\rightarrow \circled{5}$ RD & $\uline{90.17\%} \downarrow$ & $\uline{18.58\%} \downarrow$ & $\uline{80.58\%} \downarrow$ & $\uline{15.23\%} \downarrow$ \\
    \hline
    \circled{6} CNN CDA & \textbf{0.108\%} & 1.01\% & \textbf{0.16\%} & \textbf{2.02\%} \\
    \hline
   $\circled{5}\rightarrow \circled{6}$ RD & $\uline{4.42\%} \downarrow$ & $\uline{32.21\%} \downarrow$ & $\uline{20.00\%} \downarrow$ & $\uline{1.94\%} \downarrow$ \\
    \hline
    \end{tabular}%
    \vspace{-2ex}
  \end{adjustbox}

\end{table}%

\vspace{-1ex}
For the clean data adaptation, the i-vector model employs MLLR to accomplish adptation, and the CNN-based model is fine-tuned with clean data. Fine-tuning shows a more significant improvement than MLLR, especially in the clean and far-field sets. Furthermore, the system verification performance gains a limit improvement for LENA-booth and noisy set in each system. The i-vector system with MLLR and the fine-tuned CNN-based system achieved effective improvements with an averaged EER RD of +34.5\% and +51.1\% respectively.

Based on models processed by the first-step adaptation,  the cross-domain adapted i-vector system obtains a better result with a lower EER of 0.63\% in the LENA-booth set, and the CNN-based system with cross-domain adaptation achieves better results in the other three conditions, which are 0.108\%, 0.16\%, and 2.02\% in the form of EER for the clean, LENA-booth, and far-field respectively. By means of generating nearly identical distributions for all statistics in the CNN CDA, the model has a good performance on the clean set (+4.42\% RD), while it significantly improves the results on domains of LENA-booth and far-field for +32.21\% and +20\% respectively. However, the model of CNN CDA achieves limited improvement in the noisy domain. Cross-domain adaptation contributes to speaker verification performance improvement in both systems, and is more effective for the CNN-based system. The average RD of EER for the i-vector system is +13.6\%, and that for the CNN-based system is +14.6\%. Therefore, the proposed cross-domain adaptation methods have been shown to be promising/effective for new environment mismatch data scenarios.

\vspace{-2ex}
\section{Conclusions}
In this study, we explored a cross-domain method to adapt i-vector and CNN-based systems to data from new various acoustic domains. The adaptation included two steps of the clean set adaptation and cross-domain adaptation. In the clean set adaptation, an i-vector system was used with an MLLR method with clean data, and the CNN-based system employed clean data to fine-tune the model, which achieved the best result by applying fine-tuning to layer 4 in ResNet and LDE layer. Based on the first-stage adaptation, the cross-domain adaptation approach was employed to learn domain-invariant information so that the model could be greatly generalized to the other acoustic domains without degrading verification performance on the clean set. In conclusion, diverse or unknown acoustic environments which are common in field data for forensics were shown to affect speaker verification system performance, and the proposed cross-domain adaptation method was able to contribute to  verification performance improvement for multiple domains.

\label{sec:conclusion}

\vfill\pagebreak

\newpage

\bibliographystyle{IEEEtran}
\bibliography{main.bib}

\end{document}